# Side resonances and metastable excited state of NV- center in diamond

Alexander Ivanov[1] and Alexei Ivanov[1]

[1]*Immanuel Kant Baltic Federal University, Nevskogo 14, 236041 Kaliningrad, Russia.*
*aivanov023@gmail.com, aivanov@kantiana.ru*

**Abstract**

We discuss the side resonances of optically detected magnetic resonance in diamond crystal and propose the new approach to the calculation of hyperfine interaction in composed system, consisted of negatively charged nitrogen-vacancy ($NV^-$) center and nearby $^{13}C$ nuclear spin. Energy levels and spin states are obtained by new method. The base of this method is the using of complete set of commuting operators. In zero magnetic field, predicted the existence of metastable excited state.

## 1  Introduction

Control over individual electronic and nuclear spins in solid state offers promise for applications in quantum information processing (QIP) and field sensing at nanometer scale. The negatively charged nitrogen-vacancy ($NV^-$) center in diamond has recently emerged as a candidate for QIP and magnetometry [1]. The optical transitions of the $NV^-$ center allow a high degree of spin polarization at room temperature via optical pumping. By optically pumping with laser light in the green spectral range is initialized into $m_s = 0$ of its triplet ground state. Optically detected magnetic resonance (ODMR) signals enable probing the energy levels and spin states of $NV^-$ centers.

In the absence of magnetic fields, the $NV^-$ center has a magnetic resonance at a frequency of approximately 2870 MHz, which corresponds to a transition between the triplet ground-state magnetic sublevels $m_s = 0$ and $m_s = \pm 1$. Side-resonances around this central resonance have been reported in the literature. In ref. [2] attributed symmetrically located side-resonance groups separated by around 40, 260 and 300 MHz simultaneous flips of spins of the $NV^-$ center and weakly coupled a nearest-neighbor nitrogen atom in the diamond and the asymmetrically displaced side-resonances separated by around 130 MHz to hyperfine interaction of the $NV^-$ center with a nearest $^{13}C$ nuclear spin.

In this paper we propose a new method of calculation of the energy levels and spin states. The base of this method is the using of complete set of commuting operators and entangled spin states. Each state is characterized by a single set of values of commuting observables, however the properties of the spin states vary greatly in the level anti-crossing (LAC). In zero magnetic field, we predict the existence of metastable optically excited state of the $NV^-$ center with a nearest $^{13}C$ nuclear spin.

## 2  Method and results

The negatively charged $NV^-$ center in diamond consists of a substitutional nitrogen atom associated with a vacancy in an adjacent lattice site of the diamond matrix. Its ground state is a spin triplet (S=1) with an intrinsic spin quantization axis provided by the $NV^-$ center symmetry axis. We consider single $NV^-$ center associated with native $^{14}N$ isotopes (99.6 % abundance), corresponding to a nuclear spin I=1. Hyperfine coupling of the electron spin to $^{14}N$ nuclear spin at the NV⁻ center around 3 MHz [1]. The ground-state spin Hamiltonian in frequency unit reads as

$$\hat{H} = D(\hat{S}_z^2 - \hat{S}^2/3) + \gamma_e \hat{S}_z B_z \qquad (1)$$

where D≈2870 MHz is the fine structure splitting. The magnetic field is directed along the symmetry axis of NV⁻ center. Operators $\hat{S}^2$ and $\hat{S}_z$ form a complete set of commuting observables. The Hamiltonian $\hat{H}$ commutes with them, therefore, the energy E is a function of the observables S and $m_s$. This function must be a single-valued function. This means that if the observables S and $m_s$ have a certain values, the energy E automatically has a certain value. Moreover, if the observables E and S have certain values, then $m_s$ would automatically have a certain value. It is well known that this function is single-valued function everywhere except in level anti-crossing.

Level anti-crossing (LAC) appears when $B_z \approx 1028\ G$. Changes in the magnetic field strength may lead to a change in emission intensity, and this effect is particularly noticeable for an axial field of 1028 Gauss. Thus an axially aligned magnetic field swept through 1028 Gauss causes a noticeable drop in the visible emission intensity. It is generally believed (see for example [3]) that in LAC at the value of the axial magnetic field of 1028 Gauss there is a complete mixing of $m_s$=0 and $m_s = -1$ states. The population will be equally distributed between the two spin states. However, on the basis of our approach could be argued that in LAC these spin projections $m_s$ are uncertain values.

Level anti-crossing (LAC) appears also when $B_z = 0$. The two states ($m_s$=1 and $m_s = -1$) have the same energy value E=D/3 in the absence of an external magnetic field. However, on the basis of our approach could be argued that in the steady state with energy E=D/3 these spin projections $m_s$ are uncertain values.

It has been shown experimentally [4] that the relaxation rates of components of NV⁻ ODMR spectrum increase with decreasing magnetic field strength. At zero value of the magnetic field the relaxation rates of the outer and inner components merge to a common maximum value, i.e. exhibit the "zero-field relaxation resonance". In frame of our approach the reason for this effect is a sharp change in the properties of the spin states of these components in zero magnetic field.

Optical excitation of NV⁻ center is accompanied by an instantaneous change of its spin-hamiltonian and therefore its spin state is retained. The spin state is also retained in the transition with emission of a photon. It is well known that due to optical dynamics and selection rules of NV⁻ centers, the $m_s = 0$ spin state fluoresces more intensely than the $m_s = \pm 1$ spin states and also becomes nearly completely populated upon optical illumination. This effect is usually referred to as optically-induced spin polarization despite the fact that in this state the spin projection on any direction is zero. As a result, the $m_s = 0$ spin state is referred to as 'bright" and the $m_s = \pm 1$ spin states are referred to as "dark". Such spin-dependent fluorescence becomes more understandable if we take into account that in the absence of an external magnetic field in the single NV⁻ center only state with $m_s = 0$ is a state with a certain value of spin projection.

Hyperfine coupling of the electron spin of $NV^-$ center occurs to $^{13}C$ (1.1 % abundance) in the surrounding lattice. The hyperfine coupling of nearest neighbor carbons (nuclear spin I=1/2) is around 130 MHz [5]. The ground-state spin Hamiltonian in frequency unit reads as

$$\hat{H} = D(\hat{S}_z^2 - \hat{S}^2/3) + A^{\|}\hat{S}_z \hat{I}_z + A^{\perp}(\hat{S}_x \hat{I}_x + \hat{S}_y \hat{I}_y) + \gamma_e \hat{S}_z B_z + \gamma_n \hat{I}_z B_z \qquad (2)$$

where $A\|$ и $A\perp$ are the axial and non-axial magnetic hyperfine parameters of $^{13}C$ nucleus, z-axis coincides with electronic spin quantization axis.

To calculate the spectrum of the Hamiltonian we use the method that we have proposed in ref. [6]. For this purpose, the total spin operator first determines

$$\hat{\vec{J}} = \hat{\vec{S}} + \hat{\vec{I}}$$

and based on the principles of quantum theory of angular momentum will build vectors $|J, M_z\rangle$, are eigenvectors of a complete set of commuting operators $\hat{J}^2, \hat{J}_z, \hat{S}^2, \hat{I}^2$:

$$|3/2, 3/2\rangle = |1,1\rangle|1/2, 1/2\rangle$$
$$|3/2, 1/2\rangle = \sqrt{\frac{2}{3}}|1,0\rangle|1/2, 1/2\rangle + \frac{1}{\sqrt{3}}|1,1\rangle|1/2, -1/2\rangle$$
$$|3/2, -1/2\rangle = \sqrt{\frac{2}{3}}|1,0\rangle|1/2, -1/2\rangle + \frac{1}{\sqrt{3}}|1,-1\rangle|1/2, 1/2\rangle$$
$$|3/2, -3/2\rangle = |1,-1\rangle|1/2, -1/2\rangle$$

(3)

$$|1/2, 1/2\rangle = \sqrt{\frac{2}{3}}|1,1\rangle|1/2, -1/2\rangle - \frac{1}{\sqrt{3}}|1,0\rangle|1/2, 1/2\rangle$$
$$|1/2, -1/2\rangle = \frac{1}{\sqrt{3}}|1,0\rangle|1/2, -1/2\rangle - \sqrt{\frac{2}{3}}|1,-1\rangle|1/2, 1/2\rangle$$

(4)

Despite the fact that in (3) and (4) the vectors of admissible states are classified by multiplets, we note that the total spin J is not conserved, as the Hamiltonian $\hat{H}$ form (2) does not commute with the operator $\hat{J}^2$. At the same time, the Hamiltonian commutes with the projection of the total spin, the square of the electron spin and nuclear spin square:

$$[\hat{H}, \hat{J}_z] = [\hat{H}, \hat{S}^2] = [\hat{H}, \hat{I}^2] = 0 \tag{5}$$

Set of operators $\hat{H}, \hat{J}_z, \hat{S}^2, \hat{I}^2$ is also a complete set of commuting operators. This operators set have unique system of eigenvectors. Each eigenvector is characterized by a single set of commuting observables values. Consequently, the vectors of the NV⁻ center ground state characterized by the value energy E, the projection of the total spin $M_z$, electron spin S and the nuclear spin I : $|E, M_z, S, I\rangle$. Since for all these states S=1, I=1/2, then the equation for the eigenvalues and eigenvectors of the Hamiltonian (1) can be written as

$$\hat{H}\left|E_{M_z}^{(i)}\right\rangle = E_{M_z}^{(i)}\left|E_{M_z}^{(i)}\right\rangle \tag{6}$$

Where the index i is introduced in order to distinguish the states with the same value $M_z$ and different values of energy E.

On the basis of (2), (5) and (6) we can conclude that the energy E is a single-valued function of $M_z$. In other words, if the total spin projection $M_z$ has a certain value, the energy E is also well defined. It is also true the opposite proposition: if the energy E has a certain value, the total projection $M_z$ is also well defined. The condition of uniqueness is not satisfied, in particular, in the absence of an external magnetic field ($B_z = 0$).

In the absence of an external magnetic field first, we note that the states

$$\left|E_{\pm 3/2}\right\rangle = |3/2, \pm 3/2\rangle$$

are solutions of the equation (6) with eigenvalue

$$E_{3/2} = E_{-3/2} = D/3 + A^{\parallel}/2$$

Secondly, since the total spin is not conserved, the steady states with Mz=±1/2 will be represented as

$$\left|E^{(i)}_{\pm 1/2}\right\rangle = c_1^{(i)}|3/2, \pm 1/2\rangle + c_2^{(i)}|1/2, \pm 1/2\rangle \quad (7)$$

Substituting (7) into (6) and solving the resulting equation, we find the energy levels

$$E^{(1)}_{1/2} = E^{(1)}_{-1/2} = -2D/3 + V + P, \quad E^{(2)}_{1/2} = E^{(2)}_{-1/2} = -2D/3 + V - P, \quad (8)$$

where $V = D/2 - A^{\parallel}/4$, $P = \sqrt{V^2 + (A^{\perp})^2/2}$.

Thus, for the Hamiltonian (2), taking into account the laws of conservation (5) in the absence of an external magnetic field two groups of steady states are found. In particular, the states $|E_{\pm 3/2}\rangle$ and $\left|E^{(1)}_{\pm 1/2}\right\rangle$ belong to the first group. The states $\left|E^{(2)}_{\pm 1/2}\right\rangle$ belong to the second group. Radiative transitions between these states are subject to selection rule $\Delta M_z = \pm 1$. The main peak of ODMR of single NV$^-$ center occurs at a frequency $\nu_o = 2870$ MHz.

Side resonances correspond to transitions $\left|E^{(2)}_{1/2}\right\rangle \leftrightarrow \left|E^{(1)}_{-1/2}\right\rangle, \left|E^{(2)}_{-1/2}\right\rangle \leftrightarrow \left|E^{(1)}_{1/2}\right\rangle$ with frequency $\nu_h$ and transitions $\left|E^{(2)}_{1/2}\right\rangle \leftrightarrow |E_{3/2}\rangle, \left|E^{(2)}_{-1/2}\right\rangle \leftrightarrow |E_{-3/2}\rangle$ with frequency $\nu_l$, respectively. Thus, the hyperfine interaction anisotropy is shown in the structure of side resonances in diamond. Moreover, if $A^{\parallel} \neq A^{\perp}$, these side resonances are located asymmetrically relatively to the center.

In ref. [2] marked side resonances B, C, D and E. The distance between the non-symmetrical relative to the central resonance side resonances of group C is significant and equal $\Delta_C \approx 126(2)$ MHz. The reason for the formation of resonances of this group is the interaction between the NV$^-$ center and the $^{13}$C nucleus. If, based on these experimental data, take $\nu_o - \nu_l = 56$ MHz and $\nu_h - \nu_o = 70$ MHz, then for the axial and non-axial hyperfine splitting parameters, we obtain the following values $A^{\parallel} = -121$ MHz and $A^{\perp} = \pm 166$ MHz.

The authors of ref. [7] used the interaction of the electron spin NV$^-$ center with nearest $^{13}$C nuclear spin for the demonstration of quantum gate NOT and a conditional two-qubit gate. They used states of the form $|m_s\rangle|m_i\rangle$ with $m_s = \pm 1, m_i = \pm 1/2$. Note that entangled states (7) become of the form $|m_s\rangle|m_i\rangle$ only at the zero value of the parameter non-axial hyperfine splitting $A^{\perp}$. In this case, two side resonance of group C were placed symmetrically relative to the central resonance. Such arrangement of these side-resonances were not observed in the experiment. Moreover, none of the spin states of the system ($NV^-$ center with a nearest $^{13}C$) has a certain spin projection values in the absence of an external magnetic field. In other words, in this system takes place LAC for all three pairs of spin states. This is a very rare case of LAC. We can assume that such a rare total LAC leads to the formation of optically excited metastable state.

## 3 Summary

In conclusion we have studied the side resonances of optically detected magnetic resonance in diamond crystal and propose a new method of calculation of the energy levels and spin states. The base of this method is the using of complete set of commuting operators and entangled spin states. Each state is characterized by a single set of values of commuting observables, however the properties of the spin states vary greatly in the level anti-crossing (LAC). On the basis of our approach could be argued that in LAC some spin projections $m_s$ are uncertain values. In zero magnetic field, we predict the existence of metastable optically excited state of the $NV^-$ center with a nearest $^{13}C$ nuclear spin. It is obtained also the estimation of the carbon hyperfine splitting parameters in the diamond $NV^-$ center from side-resonance frequencies in the frame of this method.